\documentclass[a4paper,fleqn]{cas-sc}
\usepackage{apacite}

\usepackage[authoryear]{natbib}

\def\tsc#1{\csdef{#1}{\textsc{\lowercase{#1}}\xspace}}
\tsc{WGM}
\tsc{QE}
\tsc{EP}
\tsc{PMS}
\tsc{BEC}
\tsc{DE}

\usepackage{multicol}
\usepackage{amssymb}
\usepackage{amsmath}
\usepackage{booktabs}
\usepackage{hyperref}
\usepackage{rotating}
\usepackage{graphicx}
\usepackage{amsthm}
\usepackage{paralist}
\usepackage[labelformat=simple]{subcaption}

\usepackage{caption}

\newcommand{\eg}{e.g., }
\newcommand{\ie}{i.e. }

\newcommand{\secref}[1]{Section~\ref{#1}}
\newcommand{\figref}[1]{Figure~\ref{#1}}
\newcommand{\tableref}[1]{Table~\ref{#1}}

\begin{document}
\let\WriteBookmarks\relax
\def\floatpagepagefraction{1}
\def\textpagefraction{.001}

\shorttitle{What makes a successful rebuttal in computer science conferences?}

\shortauthors{Huang et~al.}

\title [mode = title]{What makes a successful rebuttal in computer science conferences? : A perspective on social interaction}       
\let\printorcid\relax

\author[1,2]{Junjie Huang}[style=chinese]
\ead{huangjunjie17s@ict.ac.cn}
\credit{Methodology, Data analysis, Experiments, Writing: original draft, review \& editing}

\author[3]{Win-bin Huang}[style=chinese]
\ead{huangwb@pku.edu.cn}
\credit{Methodology, Data curation, Resources, Writing: original draft, review \& editing}

\author[3]{Yi Bu}[style=chinese]
\ead{buyi@pku.edu.cn}
\credit{Methodology,  Data curation, Resources, Writing: original draft, review \& editing}

\author[1,2]{Qi Cao}[style=chinese]
\ead{caoqi@ict.ac.cn}
\credit{Methodology,  Data curation, Funding acquisition, Resources, Writing: original draft, review \& editing}

\author[1,2]{Huawei Shen}[style=chinese]
\ead{shenhuawei@ict.ac.cn}
\credit{Methodology,  Data curation, Funding acquisition, Resources, Writing: original draft, review \& editing}

\author[4]{Xueqi Cheng}[style=chinese]
\ead{cxq@ict.ac.cn}
\credit{Methodology, Funding acquisition, Resources}

\affiliation[1]{organization={Data Intelligence System Research Center, Institute of Computing Technology, Chinese Academy of Sciences},
    state={Beijing},
    country={China}}

\affiliation[2]{organization={University of Chinese Academy of Sciences},
    state={Beijing},
    country={China}}

\affiliation[3]{organization={Department of Information Management, Peking University},
    state={Beijing},
    country={China}}

\affiliation[4]{organization={CAS Key Laboratory of Network Data Science and Technology, Institute of Computing Technology, Chinese Academy of Sciences},
    state={Beijing},
    country={China}}

\begin{abstract}
With an exponential increase in submissions to top-tier Computer Science (CS) conferences, more and more conferences have introduced a rebuttal stage to the conference peer review process.
The rebuttal stage can be modeled as social interactions between authors and reviewers.
A successful rebuttal often results in an increased review score after the rebuttal stage.
In this paper, we conduct an empirical study to determine the factors contributing to a successful rebuttal using over 3,000 papers and 13,000 reviews from ICLR2022, one of the most prestigious computer science conferences.
First, we observe a significant difference in review scores before and after the rebuttal stage, which is crucial for paper acceptance.
Furthermore, we investigate factors from the reviewer's perspective using signed social network analysis.
A notable finding is the increase in balanced network structure after the rebuttal stage.
Subsequently, we evaluate several quantifiable author rebuttal strategies and their effects on review scores.
These strategies can help authors in improving their review scores.
Finally, we used machine learning models to predict rebuttal success and validated the impact of potential factors analyzed in this paper. Our experiments demonstrate that the utilization of all features proposed in this study can aid in predicting the success of the rebuttal.
In summary, this work presents a study on the impact factors of successful rebuttals from both reviewers' and authors' perspectives and lays the foundation for analyzing rebuttals with social network analysis.
\end{abstract}


\begin{keywords}
Peer Review \sep Rebuttal Analysis \sep Social Network Analysis \sep Social Interaction \sep Rebuttal Strategy \sep Rebuttal Success Prediction
\end{keywords}

\ExplSyntaxOn
\keys_set:nn { stm / mktitle } { nologo }
\ExplSyntaxOff
\maketitle

\section{Introduction}
\label{sec:intro}
Scientific peer review, the process of evaluating scientific literature by knowledgeable peers, originated in the 1700s~\citep{kronick1990peer}.
This process aims to filter out papers of low quality based on criteria such as competence, significance, novelty, and originality~\citep{brown2004peer}, thereby promoting scientific progress~\citep{shah2022challenges}.
As a cornerstone of scientific research~\citep{price2017computational}, peer review is used in almost all scientific disciplines, improves the quality of published research~\citep{jefferson2002effects}, and has achieved significant success in research evaluation~\citep{abramo2019peer}.
Peer-reviewed publications predominantly consist of conference proceedings and journals.
However, the preference for journals over conferences varies across different disciplines. Notably, studies have identified Computer Science (CS) as a discipline that values conference publications more than other academic fields~\citep{vrettas2015conferences,tomkins2017reviewer,meho2019using}.
Publishing in top-tier conferences is closely associated with a researcher's reputation, funding distribution, and tenure, among other factors~\citep{franceschet2010role,vrettas2015conferences}.
Unlike journal publications, CS conference publications have distinct timelines (i.e., specific deadlines), acceptance rate limits (around 20\%), page limits (up to 10 pages without restrictions on appendices), consistent annual meeting dates (like in May), and often require in-person discussions (such as offline meetings)\footnote{Due to COVID-19, ICLR2022 was a virtual conference; offline meetings were cancelled.}.
These features emphasize the importance of modelling social interactions in CS conferences.

Furthermore, with the surge in scientific literature, several problems have emerged in the peer review system. These are due to the scarcity of professional reviewers, the unpredictable nature of early scientific discoveries, and the difficulties in judging the potential of groundbreaking discoveries.
Issues such as the reproducibility crisis~\citep{baker2016reproducibility} and unethical behaviours under the "publish or perish" pressure~\citep{grimes2018modelling} are hindering the advancement of CS.
Peer review in CS conferences is even viewed as ineffective and arbitrary~\citep{langford2015arbitrariness,brezis2020arbitrariness}.
To mitigate the randomness in peer review and promote the publication of valuable scientific discoveries, conference organizers have implemented the rebuttal mechanism.
\href{https://openreview.net}{OpenReview} was developed to assist researchers in reviewing high-quality papers using the rebuttal mechanism~\citep{soergel2013open}.
Today, this platform is extensively used in top CS conferences like NeurIPS\footnote{https://nips.cc/}, ICML\footnote{https://icml.cc/}, and ICLR\footnote{https://iclr.cc/}.
The rebuttal mechanism offers a chance to clarify any misunderstandings of reviewers, increase the reviewer score, and enhance the possibility of paper acceptance.
Senior researchers have proposed many tips and suggestions for writing effective author responses to assist authors in navigating the rebuttal phase~\citep{noble2017ten}\footnote{https://deviparikh.medium.com/how-we-write-rebuttals-dc84742fece1}.
Yet, few studies have empirically investigated what makes a successful rebuttal by applying metrics to real-world datasets.
In this paper, we aim to bridge this gap by analysing the impact of successful rebuttals from both the reviewers' and authors' perspectives.

\begin{figure}
    \centering
    \includegraphics[width=\textwidth]{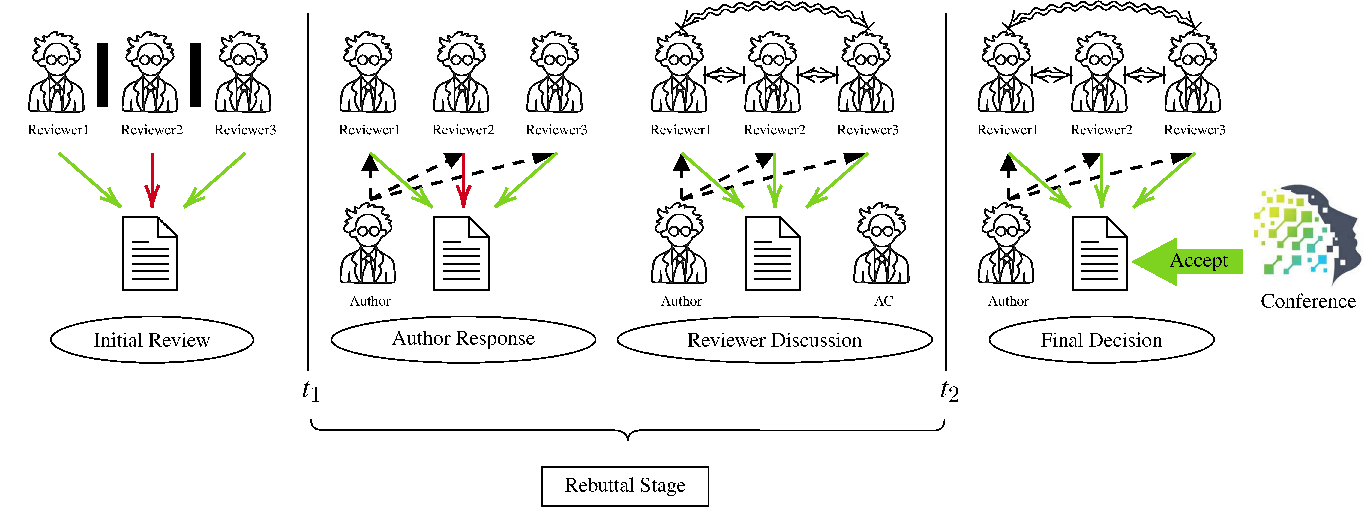}
    \caption{Illustration of peer review process in ICLR2022. ICLR2022 mainly includes four stages: initial review, author response, reviewer discussion and final decision. In this paper, we mainly focus on the rebuttal stage between $t_1$ and $t_2$.}
    \label{fig:rebuttal_process_modelling}
\end{figure}

In this paper, we select ICLR2022 as our research subject to examine the rebuttal stage in the CS conference peer review process.
ICLR2022 employs a reviewing workflow\footnote{https://iclr.cc/Conferences/2022/ReviewerGuide} that is similar to those of other major CS conferences, with a few modifications:
\begin{enumerate}
\item Initially, reviewers can bid on papers for further review based on their expertise and potential conflicts of interest.
\item Upon paper assignment, typically three or four reviewers independently assess a paper.
\item The initial review comments and scores are posted on the website for authors and other reviewers to view.
\item The rebuttal phase then commences, during which authors and reviewers can discuss the paper and the review, allowing for clarification and correction of misunderstandings.
\item After the rebuttal, reviewers can examine the authors' responses and other peer-review comments and discuss their perspectives.
\item Reviewers finalize their reviews, and Area Chairs (ACs) write meta-reviews.
\item Finally, Program Chairs (PCs) make the acceptance or rejection decisions based on reviewers' comments and meta-reviews.
\end{enumerate}
While ICLR2022 includes different roles, such as Area Chairs (ACs), Senior Area Chairs (SACs), and Program Chairs (PCs), its core process remains the same as outlined above. In this paper, we focus on this process and overlook the mechanisms of ACs, SACs, and PCs.
As shown in \figref{fig:rebuttal_process_modelling}, the core process encompasses the \textbf{initial review}, \textbf{author rebuttal}, \textbf{reviewer discussion} with reviewers and ACs, and the \textbf{final decision}.
Authors are primarily involved in paper submission, revision, and rebuttal discussions, while reviewers partake mainly in the initial independent review, rebuttal discussions with authors, and discussions with other reviewers and ACs.
This paper specifically focuses on the rebuttal stage, comprising the author rebuttal and reviewer discussion.
We examine the changes in review scores between $t_1$ and $t_2$ and investigate the factors that might influence these changes, including reviewer social pressure and author rebuttal strategy.

Based on the review process in ICLR2022, we focus on exploring the following research questions in this paper:

\begin{itemize}
\item RQ1: Does rebuttal stage matter? Is there a difference between the initial and final review results in ICLR2022? (see \secref{sec:rebuttal_results})
\item RQ2: Does  ``peer effect'' influence the score changes for reviewers? How to model it with signed social network analysis? (see \secref{sec:ssna})
\item RQ3: Are there effective strategies that authors can employ for a successful rebuttal? (see \secref{sec:strategy_analysis})
\item RQ4: Can we build machine learning models to predict whether reviewers will revise their score after rebuttal? (see \secref{sec:prediction})
\end{itemize}

The remaining sections of the paper are organized as follows:
In \secref{sec:related_work}, we provide an overview of the related works.
Then, in \secref{sec:dataset}, we present our datasets, including details about data collection, basic data description, and overall analysis.
In \secref{sec:rebuttal_results}, we analyze the rebuttal results to answer the research question on the significance of rebuttal (RQ1).
To address RQ2, we introduce Signed Social Network Analysis (SSNA) to measure the balanced motif changes in \secref{sec:ssna}.
In \secref{sec:strategy_analysis}, we conduct a strategy analysis to investigate how authors can better respond to reviewers for RQ3.
Furthermore, for RQ4, we formulate rebuttal success prediction tasks using machine learning models to examine the role of different features.
Finally, in \secref{sec:conclusion}, we present our concluding thoughts, limitations and discuss the future work.

\section{Related Work}
\label{sec:related_work}

\subsection{The Significance of Conference Proceedings in Computer Science}
Peer review is considered the "gatekeeping" process of science~\citep{siler2015measuring} and is widely adopted in almost every discipline, including Computer Science (CS).
Peer-reviewed publications include both conference proceedings and journals, but there are several differences between them, including publication quality, page limits, and timing~\citep{vardi2009conferences,fortnow2009viewpoint,vardi2010revisiting}.
For CS, some bibliometric studies demonstrate that CS has a preference for conferences over journals~\citep{vrettas2015conferences,kim2019author,meho2019using}.
The publications on prestigious conferences are related to the researcher's reputation, distribution of funds, the acceptance of research proposals, faculty positions, promotion, and tenure~\citep{franceschet2010role,vrettas2015conferences}. 

One big significance in peer review in CS is that the review of prestigious conferences (\eg ICLR, NeurIPS, ICML) usually adopts \textit{Double-blind} system, rather than \textit{Single-blind} system.
\cite{tomkins2017reviewer} suggested that using the \textit{Double-blind} system in CS conference can reduce bias in peer review~\citep{lee2013bias}.
Additionally, the rebuttal mechanism, which replaces author responses in journals, is introduced at CS conferences to reduce the arbitrariness in the peer review process~\citep{brezis2020arbitrariness}.
Compared to revision and author response in journal peer review aimed at gaining greater recognition for a paper~\citep{rigby2018journal}, the rebuttal in CS conference mainly aims to make papers accepted succeed in the competition with a low acceptance rate and massive high-quality submissions.
In this paper, we use ICLR2022 as an example to investigate the rebuttal process in CS.

\subsection{Open Peer Review Dataset}
To address the growing demand for transparency and reproducibility crisis in the peer review process, Open Peer Review (OPR) has become increasingly popular~\citep{wang2015open,ni2021influence,rigby2018journal}. 
For example, the journal Nature Communications\footnote{https://www.nature.com/ncomms/} offers authors the option to make peer reviewers' comments and their responses alongside their paper. 
Many prestigious CS conferences, such as ACL, NeurIPS, and ICLR, have joined the OPR movement based on the OpenReview system. 
Several researchers have released peer review comment datasets, including \cite{kang2018dataset}, who provided a dataset containing 14.7k paper drafts and the corresponding accept/reject decisions from top-tier CS conferences, including ACL, NeurIPS, and ICLR. 
\cite{gao2019does} presented a corpus containing over 4k reviews and 1.2k author responses from ACL2018. 
More recently, \cite{kennard2022disapere} presented DISAPERE, a labeled dataset of 20k sentences contained in 506 review-rebuttal pairs, and Dycke et al.\citep{dycke2022yes} collected a dataset for ACL Rolling Review. 
These datasets enable quantitative studies of peer review rebuttals. 
Many machine learning models have been proposed to predict whether a paper will be accepted~\citep{fernandes2022between}, perform sentiment analysis on peer review comments~\citep{ghosal2019deepsentipeer,wang2018sentiment}, and extract responses and argument pairs~\citep{cheng2020ape}. 
However, most studies focus on Natural Language Processing (NLP) rather than the social interactions between reviewers and authors.
Our study aims to identify the factors that contribute to a successful rebuttal from both authors' and reviewers' perspectives, such as reviewer social pressure and author rebuttal strategy. 
For reviewer social pressure, we primarily analyze peer effects among reviewers (see ~\secref{sec:peer_effect}).
Other social pressures include public pressure and fear of reprisals~\citep{van2010effect}.
In terms of the author's rebuttal strategy, we primarily verify the effectiveness of the tips on how to respond to peer reviewers~\citep{noble2017ten,min2022critical}.
This can enhance our understanding of the rebuttal process in peer review from a data-driven perspective.

\subsection{Peer Effect in Peer Review}
\label{sec:peer_effect}
The peer effect is a concept that describes how an individual's behavior is influenced by that of their peers~\citep{eckles2016estimating}. 
It has been extensively studied in various fields, such as education~\citep{calvo2009peer}, economics~\citep{bramoulle2020peer}, sociology, and social psychology~\citep{jencks1990social,bramoulle2020peer}. 
In peer review, a paper is usually assessed by multiple reviewers, and social influences can occur between their opinions.
\cite{gao2019does} proposed that the rebuttal in peer review is related to the opinion dynamics\citep{moussaid2013social}, which is similar to the peer effect. 
They also found that conformity bias~\citep{buechel2015opinion} plays a significant role in peer reviewing. 
However, their findings rely on machine learning models, which lack interpretability.
In this paper, we use Signed Social Network Analysis (SSNA) to model the peer effects in peer review. 
A signed network (also known as a sentiment social network~\citep{rawlings2017structural}) is a social network that contains both positive and negative links. 
The structure balance theory has been proposed to understand the structure and origin of conflicts in a social network~\citep{facchetti2011computing}.
It is a sociological theory that seeks to explain how individuals form and maintain social relationships based on the balance or imbalance of positive and negative sentiments among them~\citep{heider1946attitudes}. 
For author-referee networks, it can be modeled as a signed bipartite graph~\citep{dondio2019invisible,huang2021signed}.
\cite{dondio2019invisible} examined the influence of author-referee networks on peer review.
For the signed networks before and after the rebuttal phase,
\cite{huang2021signed} found that the ratio of balanced isomorphism (e.g., balanced signed butterfly) in signed bipartite networks increased, while the number of positive links slightly decreased.
However, obtaining such a signed bipartite graph is difficult due to privacy concerns~\citep{gao2019does}.
In this paper, we utilize SSNA to analyze the social pressure in peer review, specifically focusing on the influence between reviewers of a single paper.
Our approach offers a more interpretable method for investigating the peer effect in the peer review process.

\section{Dataset}
\label{sec:dataset}

\subsection{Data Collection}
Computer science conferences have important deadlines for authors and reviewers, such as abstract and paper submission deadlines, paper review release dates, discussion period end dates, and decision notification dates. 
For ICLR2022, the review release date was November 9, 2021, and the decision notification date was January 24, 2022. 
Detailed dates can be found on the official schedule\footnote{More detailed dates can be found at https://iclr.cc/Conferences/2022/Dates}.
Since the review results for ICLR2022, available on \href{https://openreview.net}{OpenReview}, are continually updated in real-time, there are no archived results.
(\ie The preliminary results are overwritten by the latest reviews).
Therefore, we used a web spider to crawl the website before and after the rebuttal phase at two different timestamps\footnote{The first time is 10 Nov 2021. The second time is  08 Mar 2022.}.
After data cleaning, we obtained a dataset comprising 3,338 papers, 13,021 reviews, and 37,478 replies from ICLR2022. 
ICLR2022 uses a double-blind review mechanism, which means reviewers' identities are anonymized and each is assigned a unique ID not specific to any paper.
Therefore, it is not possible to construct a signed bipartite graph as \cite{huang2021signed} did in their study.

\subsection{Data Description}
This section presents various statistics related to the decisions and presentation formats of papers submitted to ICLR2022.
In addition to the acceptance or rejection decisions, papers are also grouped according to their presentation format. 
Out of the 3,338 papers submitted, 54 were selected for Oral presentation, 176 for Spotlight presentation, and 865 for Poster presentation. 
The remaining 2,243 papers were rejected, with 1,576 papers receiving Review rejection and 667 being Desk rejection/Withdrawn cases.

Similar to journal peer review, the ICLR2022 organizers may add more reviewers to papers if they receive insufficient or widely varying reviews during the preliminary review. 
In our dataset, 139 papers had additional reviewers assigned to them, with 123 papers receiving one additional reviewer, 13 papers receiving two additional reviewers, and three papers receiving three additional reviewers. 
Interestingly, 73 out of the 139 papers (52.5\%) experienced an increase in their average score after the additional reviews.
The proportion of papers with increased scores after additional reviews exceeded the overall percentage of papers that had an increase in score post-rebuttal (43.59\%).

In ICLR2022, reviewers were asked to provide a recommendation score to express their opinions on the papers under review. 
Reviewers chose their scores from a range of \{1, 3, 5, 6, 8, 10\} (\textit{1: strong reject; 3: reject, not good enough; 5:  marginally below the acceptance; 6:  marginally above the acceptance; 8: accept, good paper; 10: strong accept, should be highlighted at the conference})

Of the 13,021 reviews collected in our dataset, 304 reviews assigned a score of \textit{strong reject}, 3,167 of \textit{reject}, 3,710 of \textit{marginally below the acceptance}, 3,698 of \textit{marginally above the acceptance}, 2,088 of \textit{accept}, and 54 of \textit{strong accept}.
To facilitate analysis, we treated scores of 6, 8, and 10 as positive and scores of 1, 3, and 5 as negative. 
The resulting positive ratio stood at 0.449, surpassing the positive ratios reported in some related works~\citep{gao2019does,huang2021signed} (refer to \secref{sec:ssna}).

\subsection{Overall Analysis}
In this section, we provide an overall analysis of both papers and reviews in our dataset. 
This analysis forms the foundation for our study and enhances our understanding of the ICLR2022 peer review process.

\subsubsection{Paper Analysis}
\begin{table}

\centering
\caption{Statistics of the ICLR2022 papers for different decision groups. (Mean$\pm$SD)}
\label{tab:paper-analysis}
\resizebox{\textwidth}{!}{
\begin{tabular}{lrrrrrrrr}
\toprule

& Oral    & Spotlight & Poster & \begin{tabular}[c]{@{}c@{}} Review \\Reject \end{tabular} & \begin{tabular}[c]{@{}c@{}}Desk Reject \\Withdrawn \end{tabular}  & Accept & Reject  & p-value  \\
\midrule
Number                   & 54     & 176   & 865    & 1576   & 667    & 1095   & 2243   &        \\
\# Authors                 & 5.33$\pm$2.96  & 4.77$\pm$3.38  & 4.77$\pm$2.11  & 4.27$\pm$1.98  & 4.26$\pm$1.92  & 4.80$\pm$2.40  & 4.27$\pm$1.96  & $<1e^{-3}$   \\
\# Title Word Count            & 8.28$\pm$2.53  & 7.98$\pm$2.51  & 8.32$\pm$2.75  & 8.38$\pm$2.74  & 8.25$\pm$2.78  & 8.27$\pm$2.70  & 8.34$\pm$2.75  & 0.46   \\
\# Abstract Word Count         & 200.13$\pm$42.27 &	205.63$\pm$51.90 &	200.29$\pm$46.83 &	199.92$\pm$51.73 &	203.03$\pm$49.89 &	201.14$\pm$47.47 &	200.84$\pm$51.20 &	0.87 \\
\# Keyword              & 3.72$\pm$2.24  & 3.66$\pm$2.39  & 3.47$\pm$1.94  & 3.34$\pm$2.03  & 3.0$\pm$2.06  & 3.52$\pm$2.04  & 3.24$\pm$2.04  & $<1e^{-3}$   \\
\# Paper Page               & 26.48$\pm$11.20 & 25.92$\pm$16.62 & 23.78$\pm$10.21 & 19.30$\pm$7.49  & 16.62$\pm$5.80  & 24.26$\pm$11.55 & 18.50$\pm$7.13  & $<1e^{-3}$  \\
\# Revision               & 5.22$\pm$1.76  & 5.18$\pm$1.89  & 5.46$\pm$2.42  & 3.44$\pm$1.89  & 2.38$\pm$1.08  & 5.41$\pm$2.31  & 3.12$\pm$1.76  & $<1e^{-3}$  \\
Has One-sentence Summary     & 0.74$\pm$0.44  & 0.83$\pm$0.38  & 0.79$\pm$0.41  & 0.75$\pm$0.43 & 0.73$\pm$0.44  & 0.80$\pm$0.40  &   0.75$\pm$0.44 & $<1e^{-3}$  \\
Has Supplementary Material    & 0.48$\pm$0.50  & 0.44$\pm$0.50  & 0.48$\pm$0.50  & 0.46$\pm$0.50 & 0.40$\pm$0.49  & 0.47$\pm$0.50  &   0.45$\pm$0.50 & 0.18   \\
Has Skillful Author  & 0.83$\pm$0.38  & 0.84$\pm$0.37  & 0.79$\pm$0.41  & 0.63$\pm$0.48 & 0.56$\pm$0.50  & 0.80$\pm$0.40  &   0.60$\pm$0.49 & $<1e^{-3}$ \\
Has Best Computer Science Scientists & 0.35$\pm$0.48 & 0.24$\pm$0.43 & 0.26$\pm$0.44 & 0.17$\pm$0.38 & 0.19$\pm$0.39 & 0.26$\pm$0.44 & 0.18$\pm$0.38 & $<1e^{-3}$ \\
\bottomrule
\end{tabular}
}
\end{table}

\begin{table}

\begin{center}
\caption{Statistics of the ICLR2022 reviews for different decision groups (Mean$\pm$SD).}
\label{tab:review-analysis}
\resizebox{\textwidth}{!}{
\begin{tabular}{lrrrrrrrrr}
\toprule	
& \begin{tabular}[c]{@{}c@{}} Strong  \\Reject \end{tabular}     
& Reject 
 & \begin{tabular}[c]{@{}c@{}} Below  \\Acceptance \end{tabular}  
& \begin{tabular}[c]{@{}c@{}} Above  \\Acceptance \end{tabular} 
& Accept 
& \begin{tabular}[c]{@{}c@{}} Strong  \\Accept \end{tabular}
& Positive
& Negative  
& p-value  \\
\midrule
Number   & 304 & 3,167 & 3,710 & 3,698 & 2,088 & 54 & 5,840 & 7,181         \\
Correctness	 &2.00$\pm$0.83 &	2.62$\pm$0.73 &	3.04$\pm$0.62 &3.31$\pm$0.57	&3.62$\pm$0.51 &3.93$\pm$0.26 &3.43$\pm$0.57 &2.81$\pm$0.73 & $<1e^{-3}$\\
Technical Novelty and Significance
	 &1.36$\pm$0.56	&1.99$\pm$0.57 &2.40$\pm$0.58 &2.75$\pm$0.58 &3.14$\pm$0.58 &3.59$\pm$0.50	&2.90$\pm$0.61	&2.18$\pm$0.63 &$<1e^{-3}$ \\
Empirical Novelty and Significance &1.32$\pm$0.49 &1.97$\pm$0.55 &2.35$\pm$0.58 &2.74$\pm$0.60 &3.13$\pm$0.60 &3.51$\pm$0.62	&2.89$\pm$0.63	&2.14$\pm$0.62 &$<1e^{-3}$ \\
Confidence 	 &4.32$\pm$0.75 	&3.92$\pm$0.73 &3.65$\pm$0.74 &3.54$\pm$0.76 &3.65$\pm$0.74 &4.02$\pm$0.71 &3.58$\pm$0.76 &3.80$\pm$0.76 &$<1e^{-3}$\\

Positive Score  & 0.03$\pm$0.13 &0.09$\pm$0.23 &0.18$\pm$0.33 &0.36$\pm$0.41 &0.57$\pm$0.43 &0.78$\pm$0.37 &0.44$\pm$0.43 &0.13$\pm$0.29 & $<1e^{-3}$
\\

\# Reviewer Replay & 1.25$\pm$0.55 &1.30$\pm$0.64 &1.40$\pm$0.66 &1.54$\pm$0.71 &1.57$\pm$0.68 &1.61$\pm$0.74 &1.55$\pm$0.70 &1.35$\pm$0.65 &$<1e^{-3}$ \\

\# Review Word Count	 &505.96$\pm$404.92	&562.99$\pm$354.31	&514.54$\pm$294.72	&492.94$\pm$288.58	&514.90$\pm$311.34	&643.07$\pm$511.92	&502.18$\pm$300.08	&535.55$\pm$328.24	&$<1e^{-3}$ \\
\bottomrule
\end{tabular}
}
\end{center}
\end{table}

In this subsection, we analyze the difference between accepted papers and rejected papers.
First, we analyze the number of authors in each group. 
Subsequently, we compute the word count of the paper's meta information, including the title, abstract, and keywords.
Additionally, we ascertain whether the paper includes a one-sentence summary and whether supplementary material was uploaded, as per the options provided by ICLR2022.
In addition, we investigate whether having skillful authors impacts the acceptance of a paper. 
We define a binary variable to indicate whether any author of a paper has previously published in an ICLR conference or is listed among the top Computer Science Scientists. (It is based on a scholar's D-index (Discipline H-index~\citep{hirsch2005index}))\footnote{https://research.com/scientists-rankings/computer-science}.
Additionally, we quantify the number of revisions and pages per paper.
Finally, we perform a two-sample t-test between accepted and rejected papers and report the corresponding p-value.
The results are listed in \tableref{tab:paper-analysis}.
From \tableref{tab:paper-analysis}, we can infer that:
\begin{inparaenum}[(1)]
    \item The number of authors for accepted papers is significantly higher than that of rejected papers (4.80 $>$ 4.27). 
    Papers accepted for Oral presentations have the highest average author count (5.33), while those that wereDesk Reject/Withdrawn have the lowest average author count (4.26).
    \item The length of the title and abstract does not show any significant difference between accepted and rejected papers. 
    Therefore, it appears that the length of the title and abstract does not significantly impact the paper's acceptance.
     \item There are significant differences between accepted and rejected papers in the number of keywords and whether the one-sentence summary is completed. 
     Our analysis suggests that keywords and one-sentence summaries might play a more significant role in aligning with reviewers' bids than titles and abstracts do.
    \item Although the ratio of material uploaded is slightly higher for accepted papers (0.47 compared to 0.45), there is no significant difference in the upload of supplementary material between accepted and rejected papers.
    
    \item The proportion of accepted papers with skilled authors is 0.80, which is significantly higher than that of rejected papers (0.60). 
    This finding demonstrates that seasoned authors have a higher probability of paper acceptance compared to first-time authors.
    Additionally, accepted papers include a higher proportion of top-rated Computer Science Scientists as authors, indicative of the Matthew effect~\citep{merton1968matthew} prevalent in CS conferences.
    \item Accepted papers have significantly more pages and revisions than rejected papers. 
    This indicates that a higher page count (typically including appendices) and more revisions could potentially enhance the probability of a paper's acceptance.
\end{inparaenum}

\subsubsection{Review Analysis}

In this subsection, we examine the differences between positive and negative reviews.
Initially, we analyze the detailed scoring metrics of each review across various groups to understand their correlation with recommendation scores.
Next, we investigate reviewer rebuttal activity by counting the number of replies.
Subsequently, we employ sentiment analysis to discern if a review adopts a more negative tone when the paper is considered inadequate (i.e., when the recommendation score is less than 6.0).
We use a fine-tuned DistilBERT model~\citep{sanh2019distilbert} as the tool to analyze the review texts data and calculate a positive sentiment score, ranging from 0 to 1, where a value less than 0.5 indicates negative sentiment.
Furthermore, we calculate the word count of the review text to compare the length of reviews across different groups.
Finally, we perform a two-sample t-test \footnote{In cases where the variances of the two groups are unequal, we utilize Welch's t-test.} to evaluate the differences between positive and negative reviews, and we report the corresponding p-value.
The results are listed in \tableref{tab:review-analysis}.
From \tableref{tab:review-analysis}, we can find that:
\begin{inparaenum}[(1)]
    \item Positive reviews (i.e., reviews for accepted papers) score significantly higher in aspects such as correctness, novelty, and significance compared to negative reviews (i.e., reviews for rejected papers), indicating that reviewers exhibit greater confidence in identifying flaws.
    \item Reviewers demonstrate more active engagement in discussions for positive reviews (i.e., 1.55 compared to 1.35), suggesting that positive reviews tend to foster more active discussions.
    \item 
    Reviews favoring acceptance yielded significantly higher positive sentiment scores compared to reviews advocating rejection (i.e., 0.44 compared to 0.13), suggesting that recommendation scores correspond to the positive or negative sentiment expressed in the review text.
    \item Despite negative reviews exhibiting a significantly higher word count than positive reviews, the reviews expressing strong acceptance recorded the highest word counts.
\end{inparaenum}

\section{Rebuttal Results}
\label{sec:rebuttal_results}

\begin{table}
\centering
\caption{Statistics of different types of reviews and papers.}
\label{tab:rebuttal-analysis}
\begin{tabular}{lcccc} 
\toprule
& \#Review	&\#Paper	&\%Accept & $\Delta$ \\
\midrule
KEEP  & 10,374                      & 1,727                      & 13.26\%                      & 4.52 $\rightarrow$ 4.52  \\ 
\midrule
INC   & 2,310                       & 1,444                      & 58.38\%                      & 5.37 $\rightarrow$ 6.10  \\ 
\midrule
DEC   & 179                        & 167                       & 13.77\%                      & 5.17 $\rightarrow$ 4.74  \\ 
\midrule
Total & 12,863 & 3,338 & 32.80\% & 4.92 $\rightarrow$ 5.22  \\
\bottomrule
\end{tabular}
\end{table}

In this section, we address our first research question (RQ1): How impactful is the rebuttal stage?
\tableref{tab:rebuttal-analysis} displays the number of reviews that increase (INC), decrease (DEC), or maintain (KEEP) their recommendation scores after the rebuttal stage (\#Review).
Additionally, we present the average paper scores (\#Paper), changes in average paper scores ($\Delta$), and acceptance percentages in \tableref{tab:rebuttal-analysis}.
The average scores of 1,444 papers show an increase, resulting in an acceptance rate of approximately 58.38\%, significantly higher than the acceptance rates of the 167 papers with decreased scores (13.77\%) and the 1,727 papers with unchanged scores (13.26\%).
While 43.25\% (1,444/3,338) of papers experienced an increase in scores, only 17.95\% (2,310/12,863) of reviews displayed a similar increase
Changes in the scores significantly impact paper acceptance.
Subsequently, we delve deeper into the analysis of score changes, considering both the perspectives of the paper and the review.

\subsection{Paper Perspective}

\begin{figure}
\centering
\begin{subfigure}[t]{0.8\textwidth}
    \includegraphics[width=\linewidth]{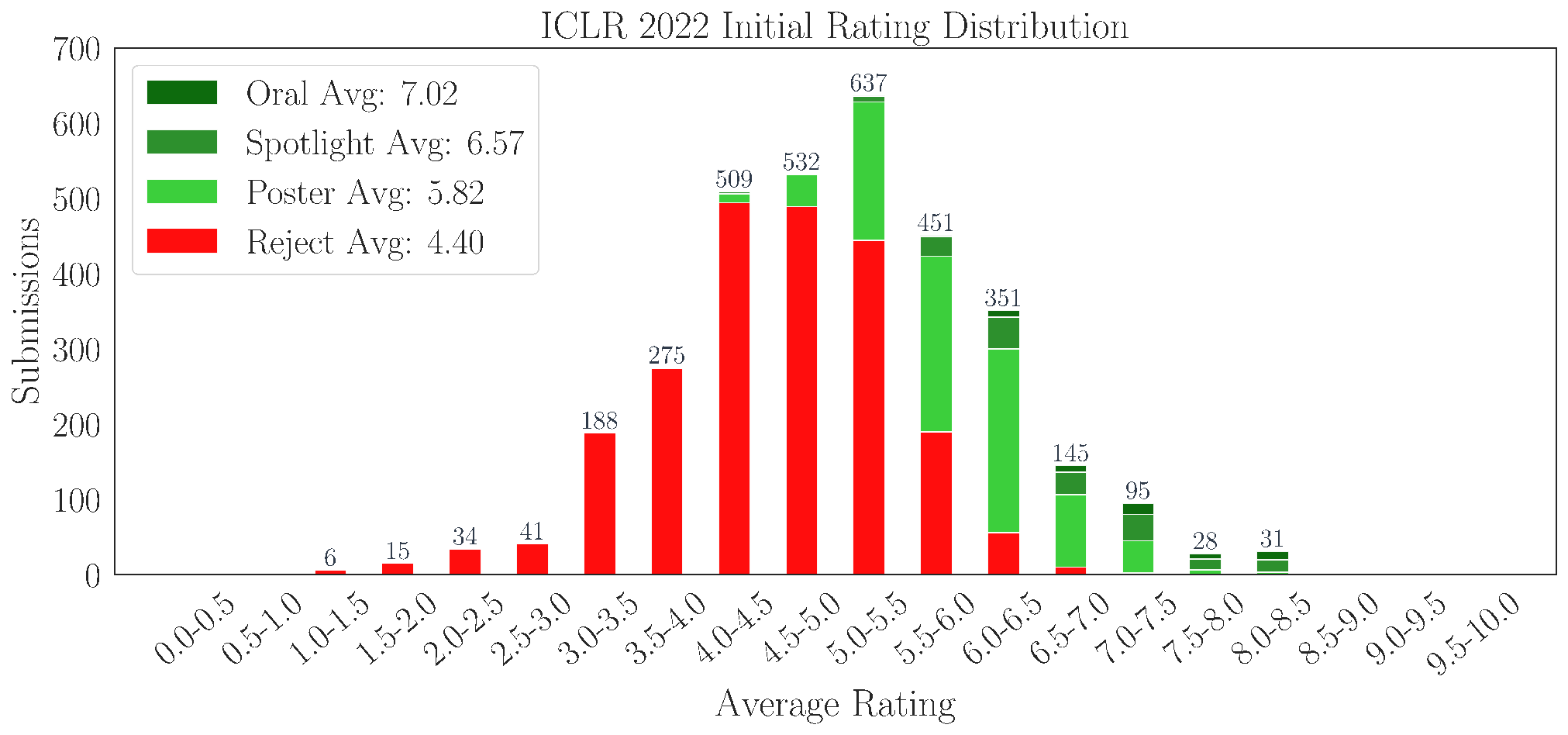}
    \caption{Initial paper average score.}
    \label{fig:inital_score_distrbution}

\end{subfigure}
     \vfill
\begin{subfigure}[t]{0.8\textwidth}

\centering
    \includegraphics[width=\linewidth]{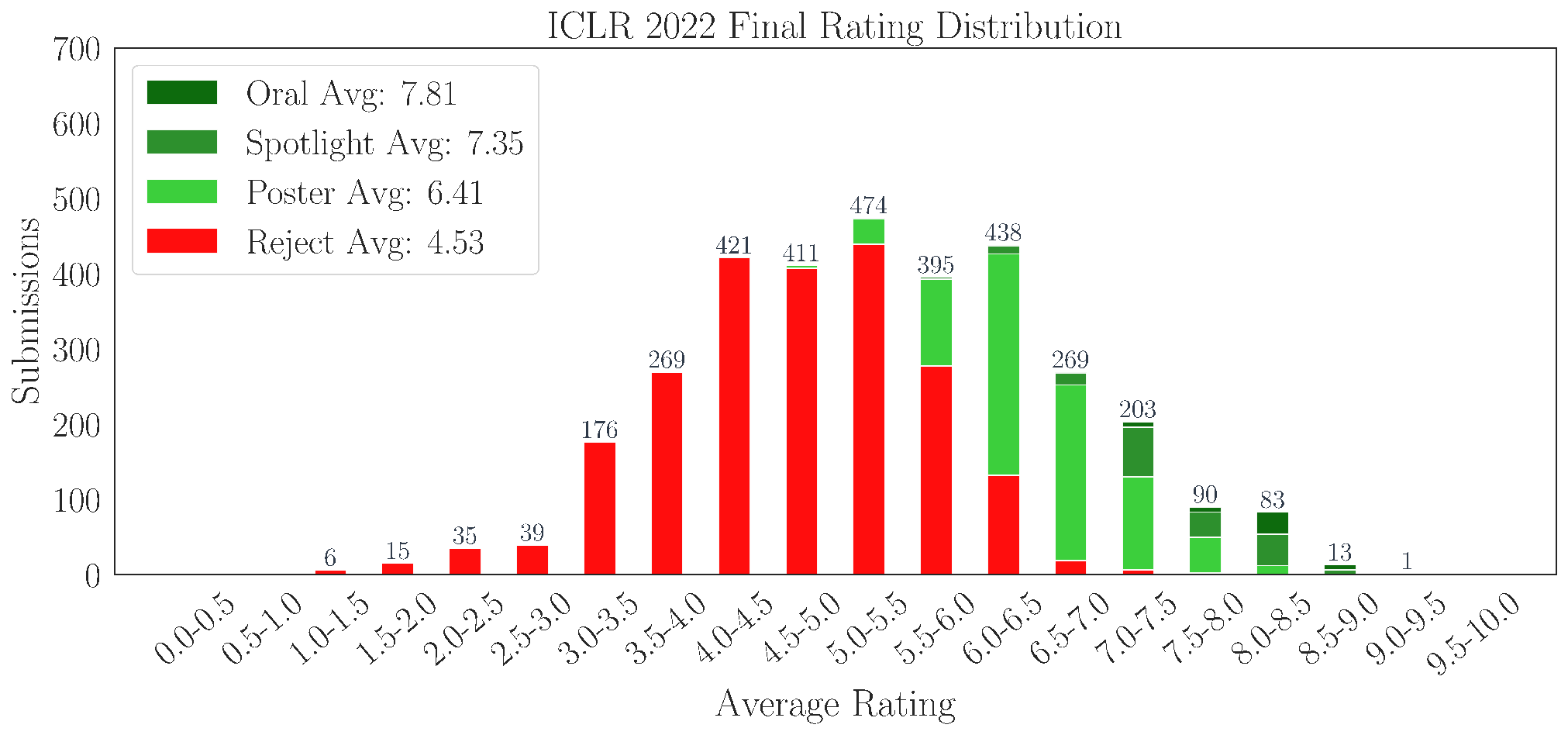}
    \caption{Final paper average score.}
    \label{fig:final_score_distrbution}
\end{subfigure}

\caption{The distribution of  paper average score, including both initial rating and final rating.}
\end{figure}

\begin{figure}
     \centering
     \begin{subfigure}[b]{0.4\textwidth}
         \centering
         \includegraphics[width=\linewidth]{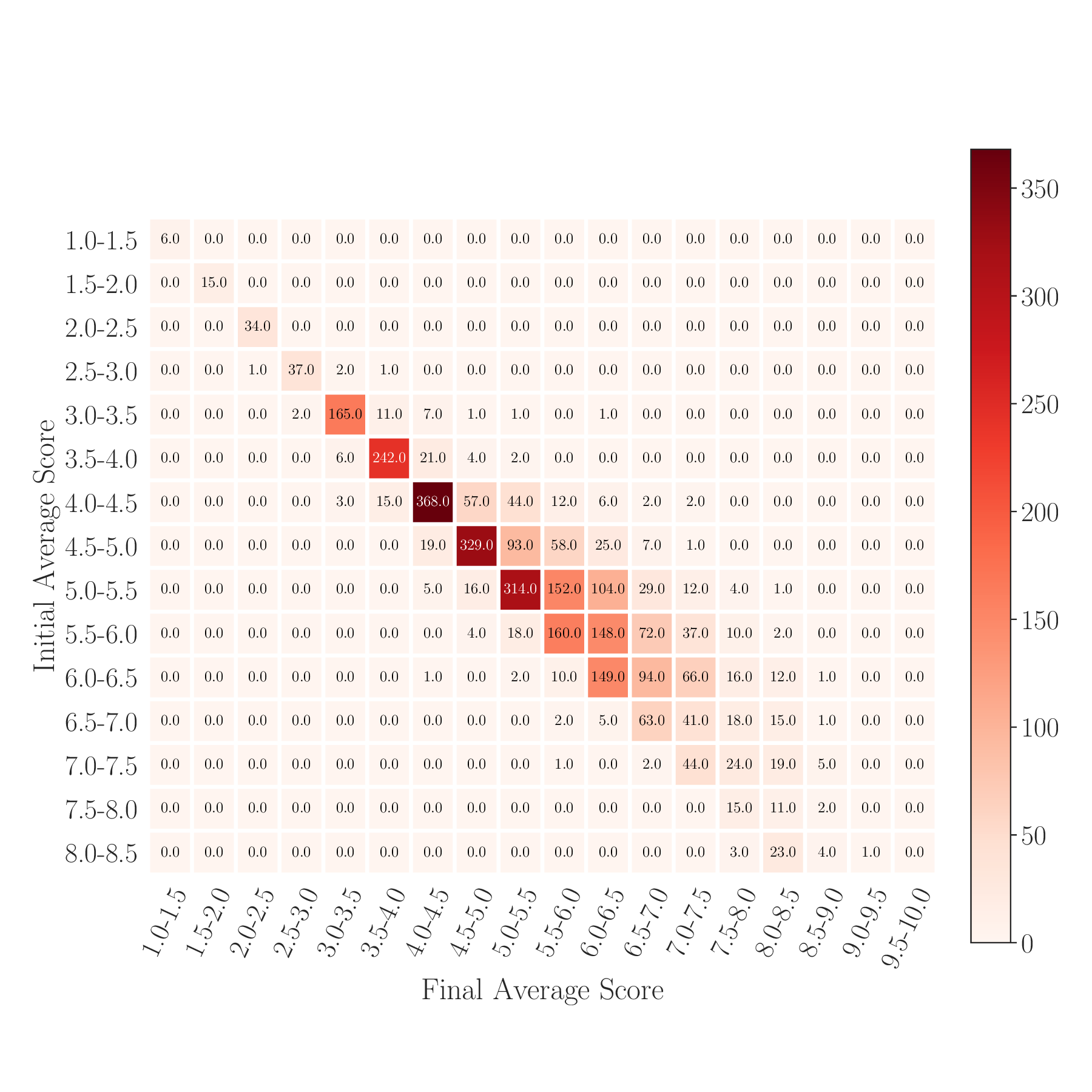}
        \caption{The heatmap.}
        \label{fig:heapmap_paper}
     \end{subfigure}
     \hfill
     \begin{subfigure}[b]{0.59\textwidth}
         \centering
         \includegraphics[width=\linewidth]{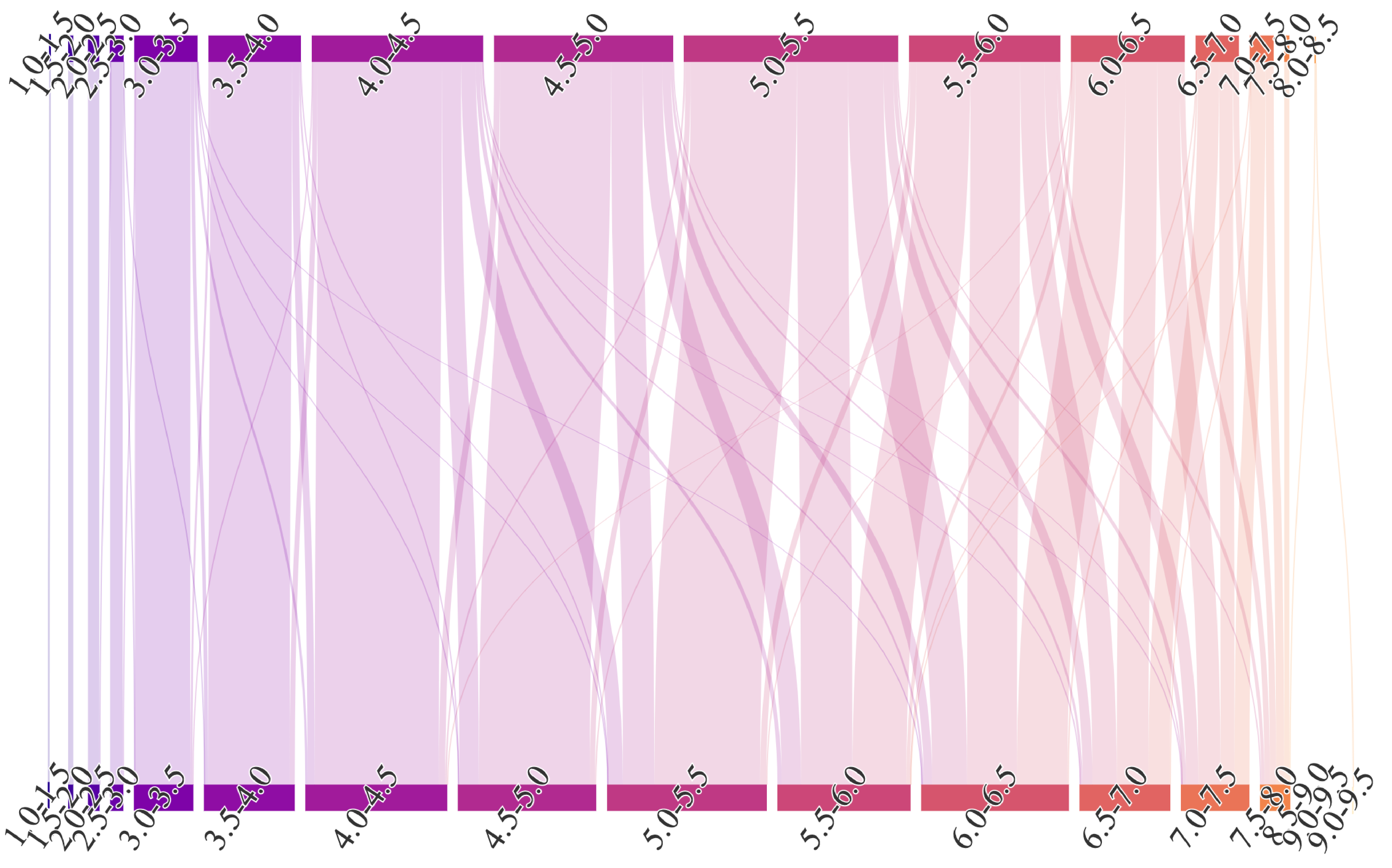}
         \caption{The Sankey diagram.}
         \label{fig:sankey_paper}
     \end{subfigure}
     \caption{The heatmap and Sankey diagram of paper average score change.}
\end{figure}

We display the average score distribution for different acceptance groups in \figref{fig:inital_score_distrbution} and \figref{fig:final_score_distrbution}.
From \figref{fig:inital_score_distrbution} and \figref{fig:final_score_distrbution}, we observe that the original average score cut-off for a paper to be accepted is between 5.0 and 6.0.
When the initial score is below 5.0, the probability of acceptance is low.
For the final average score, the cut-off is 6.0.
This demonstrates that the rebuttal stage is crucial, and most accepted papers experience an increase in score after rebuttal.
To illustrate the increase in average scores, we use the Sankey diagram in \figref{fig:sankey_paper} and categorical heatmap in \figref{fig:heapmap_paper} to display the paper changes.
From \figref{fig:sankey_paper} and \figref{fig:heapmap_paper}, we find that while a smaller percentage of papers in all score groups show a decrease (DEC) in average scores, most papers maintain (KEEP) or increase (INC) in average scores after rebuttal.
In particular, we observe a higher percentage of increases (INC) when the average scores of papers are above 6.0 compared to when they are maintained (KEEP) or decreased (DEC).
The results mentioned above reflect the peer review process requires accepted papers to meet a minimum standard of quality.
Besides, it indicates that the rebuttal stage primarily benefits borderline papers, helping them gain acceptance.

\subsection{Review Perspective}

\begin{figure}
     \centering
     \begin{subfigure}[b]{0.4\textwidth}
         \centering
         \includegraphics[width=\linewidth]{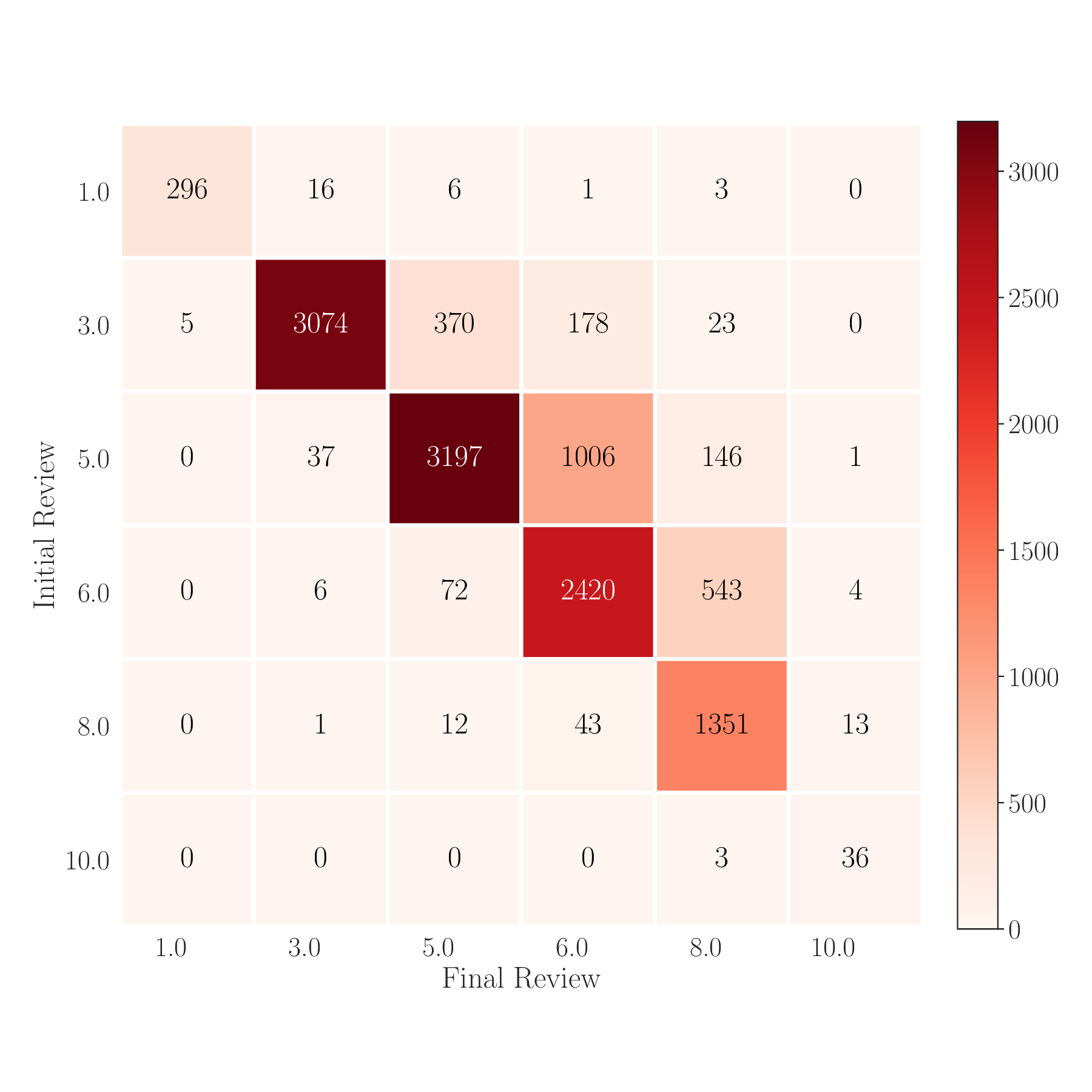}
         \caption{The heatmap}
         \label{fig:heapmap_review}
     \end{subfigure}
     \hfill
     \begin{subfigure}[b]{0.59\textwidth}
         \centering
         \includegraphics[width=\linewidth]{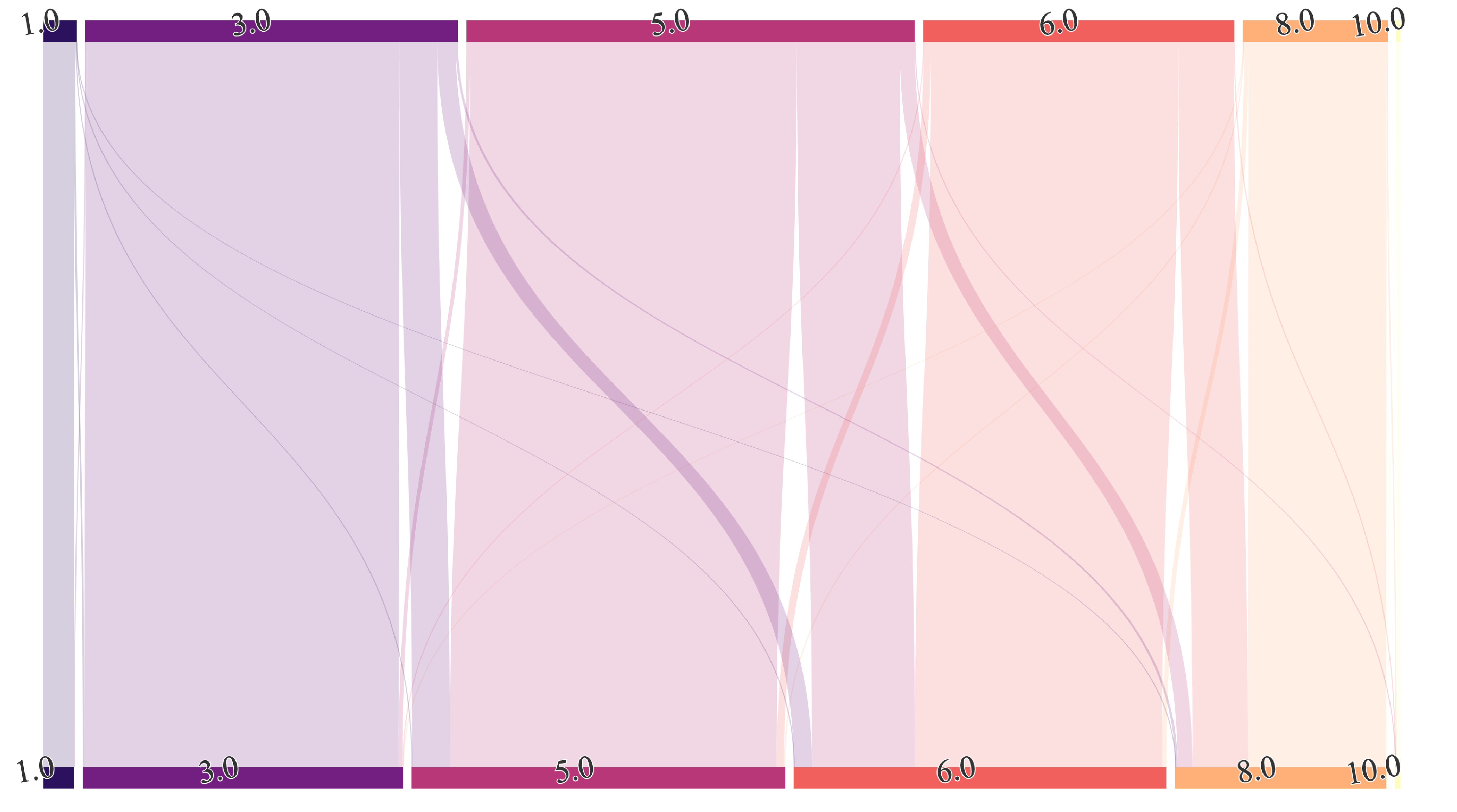}
         \caption{The Sankey diagram}
         \label{fig:sankey_review}
     \end{subfigure}
     \caption{The heatmap and Sankey diagram of review score change.}
\end{figure}

\begin{figure}
    \centering
    \includegraphics[width=0.7\linewidth]{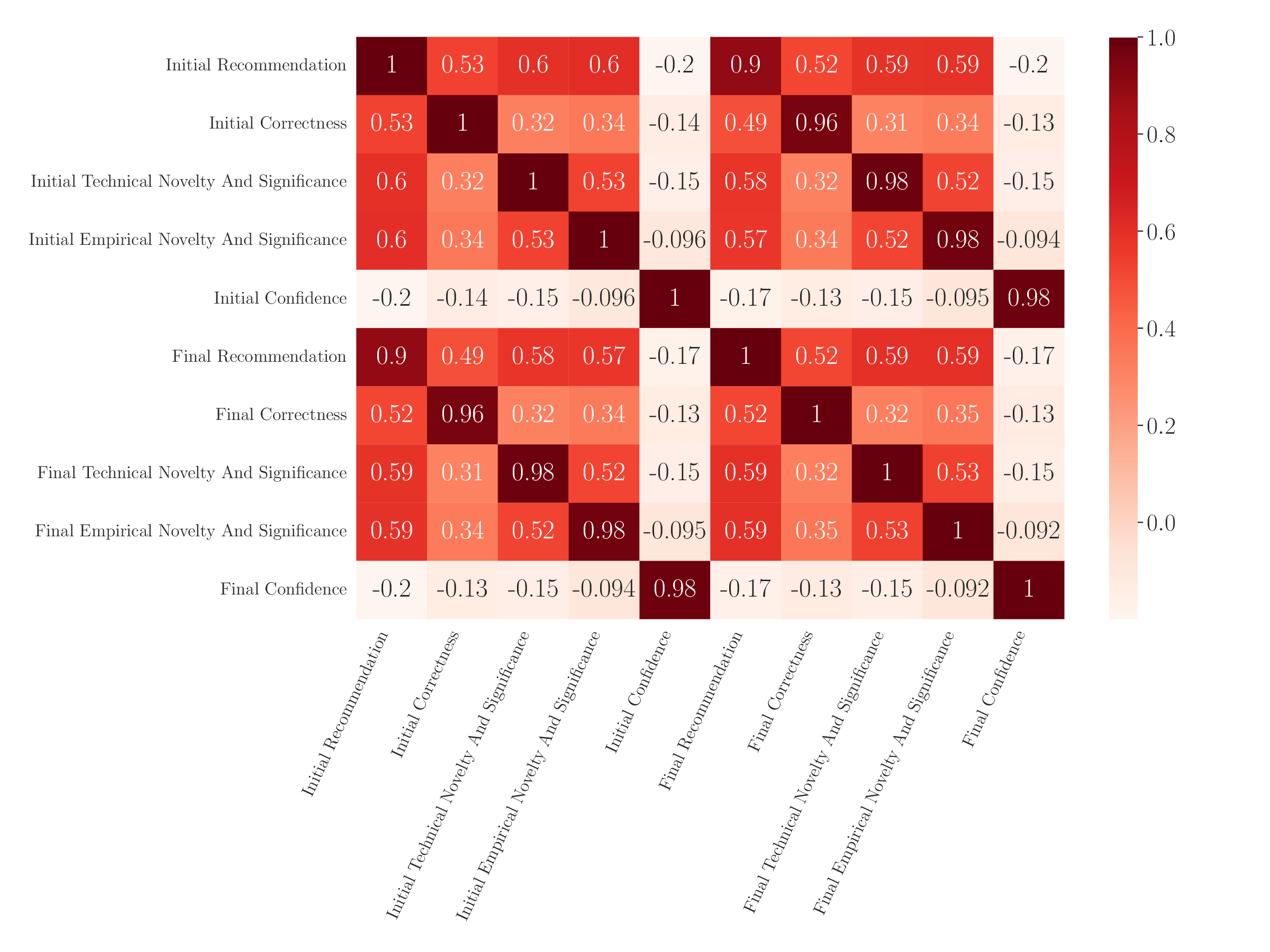}
    \caption{The detailed score correlation matrix before and after rebuttal.}
    \label{fig:detailed_score_corr}
\end{figure}

We display the changes in review scores using the Sankey diagram in \figref{fig:sankey_review} and categorical heatmap in \figref{fig:heapmap_review}.
From Figures \ref{fig:sankey_review} and \ref{fig:heapmap_review}, we observe that most reviewers do not change their initial review scores. The majority of score increases (INC) occur at the borderline (\ie around 5.0), which may help the paper gain acceptance.
For high scores (\ie 8.0), the percentage of decreases (DEC) is larger than increases (INC), but both are smaller than maintenance (KEEP).
This finding is consistent with the results in \cite{gao2019does}.
\figref{fig:detailed_score_corr} shows the correlation of different scores before and after rebuttal.
We find that the recommendation score is positively correlated with other detailed scores but negatively correlated with confidence.
Furthermore, the final review scores are positively correlated with the initial review scores.

In conclusion, changes in scores before and after rebuttal do exist and have a more significant impact on borderline papers. 
In top-tier CS conferences, the limitation of acceptance rates makes borderline papers highly competitive; authors strive to increase the recommendation score to ensure their papers are accepted.

\section{Signed Social Network Analysis}
\label{sec:ssna}

In peer review, reviewers may be influenced by the behavior of their peers. 
For example, reviewers might change their decision to reject a paper based on accepted recommendations from other reviewers for the same paper. 
\cite{gao2019does} point out that "peer pressure" is the most important factor for score changes. They use all peer review scores for a given submission to build features, including before-rebuttal scores, the statistics of other peer reviews' scores, and statistics of all peer reviews' scores (\ie max/min/mean/median/std). 
Although these metrics are considered the most important indicators for rebuttal analysis~\citep{gao2019does}, we argue that these metrics cannot reflect the number of reviewers and lack interpretability of the rebuttal results. 

\begin{figure}
    \centering
    \includegraphics[width=0.6\linewidth]{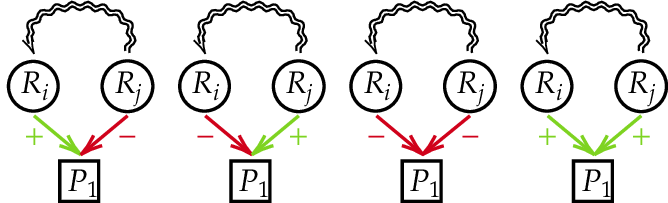}
    \caption{The illustration of signed motifs in peer reviews. we define the first two motifs as unbalanced, and the last two motifs as balanced. }
    \label{fig:signed_motifs}
\end{figure}

In this paper, we propose using Signed Social Network Analysis (SSNA) to analyze the rebuttal process. 
SSNA is based on balance theory developed by Fritz Heider in the 1950s~\citep{heider1946attitudes}.
It is a psychological theory that aims to explain how individuals strive to maintain cognitive consistency in their attitudes and perceptions about themselves, others, and objects in their social environment. 
According to this theory, people are more comfortable with balanced relationships and feel an inner tension when they experience imbalance. 
In our case, we first define the following four signed motifs for a link from reviewer $R_i$ to paper $P_1$ considering reviewer $R_j$. 
According to balance theory, we define the first two motifs in \figref{fig:signed_motifs} as unbalanced, and the last two motifs as balanced. 
The first two motifs mean that reviewer $R_j$ will "negatively" affect $R_i$, which may cause $R_i$ to change their score. 
For example, in the first motif, $R_i$ gives a positive score to paper $P_1$. 
But when $R_j$ gives a negative score to $P_1$, it will cause "peer pressure" on $R_i$. 
When more reviewers exert this pressure, it is very likely that $R_i$ will revise their score.

To test whether the above theories hold in the peer-reviewed rebuttal scenario, we perform validation measurements on three top computer science conference datasets. 
In addition to the dataset of ICLR2022 in this paper, we also include the data of ACL2018~\citep{gao2019does} and a top computer science conference (TCSC)~\citep{huang2021signed}. 
First, we assign review signs based on different scoring scales. In ACL2018, the conference adopts a 6-point scale (\textit{1: clear reject, ..., 4: worth accepting, 5: clear accept, 6: award-level}), and the sign interval is 3. In ICLR2022, the conference uses {1, 3, 5, 6, 8, 10}, and the sign interval is 5.0. 
For TCSC, the dataset is provided with signs. Then, we report the ratio of positive links before and after rebuttal. 
Next, we count the total number of balanced and unbalanced motifs from different conferences. 
For each reviewer, we extract the number of distinct motifs in  \figref{fig:signed_motifs}. 
Additionally, we compute the proportion of balanced motifs in each paper (\eg, the result of [-1, 1, 1] is 33.3\%), and we perform a paired t-test before and after the rebuttal stage to verify whether the changes are significant.

\begin{table}
\centering
\caption{The signed network analysis on three top computer science conference datasets. ($^{*}$ means that the p-value < 1e-3 for paired t-test)}
\label{tab:sign_analysis}
\begin{tabular}{lrrrrrr}
\toprule
 & \multicolumn{2}{c}{ICLR2022} & \multicolumn{2}{c}{ACL2018} & \multicolumn{2}{c}{TCSC}  \\
\midrule
               & Before   & After  & Before  & After  & Before          & After   \\
\# Links      & 12,863	& 13,021	 &3,875	 &4,054	& 1,170	& 1,170 \\
\% Positive Links &35.0 &44.9($\uparrow$) &43.0 &42.3($\downarrow$) &40.3 &39.7($\downarrow$) \\
\# Balanced motifs      & 11,720    & 13,329($\uparrow$)  & 2,324    & 2,679($\uparrow$)   & 1,002            & 1,134($\uparrow$)    \\
\# Unbalaned  motifs    & 7,208     & 6,087($\downarrow$)  & 1,098    & 1,044($\downarrow$)   & 690             & 558($\downarrow$)    \\ %
\% Averaged Positive Ratio & 35.2  &45.1$^{*}$($\uparrow$) &43.0 &41.9($\downarrow$) &40.2 &39.7($\downarrow$)    \\
\% Averaged Balanced Ratio & 61.7  &68.7$^{*}$($\uparrow$) & 61.8  & 69.7$^{*}$($\uparrow$)  & 58.5  & 66.1$^{*}$($\uparrow$)    \\
\bottomrule
\end{tabular}
\end{table}

We report the average score of all papers in \tableref{tab:sign_analysis}.
From \tableref{tab:sign_analysis}, we observe that:
\begin{inparaenum}[(1)]
\item The ratio of positive links is below 50\% across all three datasets. 
We hypothesize that this is due to the need to control acceptance rates in top computer science conferences (\eg, 24.9\% in ACL2018).
\item In all three datasets, the number of balanced motifs and the proportion of balanced motifs per paper increase after rebuttal and the unbalanced ones decrease (\eg 11,720 $\rightarrow$ 13,329($\uparrow$) and 7,208 $\rightarrow$ 6,087($\downarrow$)).
The proportion of balanced motifs per paper significantly increases after the rebuttal stage.
This finding validates the balance theory in this scenario, suggesting that reviewers tend to reduce conflict in review comments, which helps the review results reach an agreement.
\item Another interesting observation is that, unlike ICLR2022, the negative link ratio of ACL2018 and TCSC increases after rebuttal.
We speculate that this is because review comments will eventually be made public in ICLR2022, but not in ACL2018 and TCSC.
The public social peer~\citep{lee2013bias} pressure may be the primary reason for these surprising findings.
\end{inparaenum}

\section{Strategy Analysis}
\label{sec:strategy_analysis}

\begin{table}
\centering
\caption{Results of rebuttal strategy analysis (Mean$\pm$SD).}
\label{tab:strategy_analysis}
\scalebox{0.9}{
\begin{tabular}{rrcccc}
\toprule
Strategy & Metrics & $G_0$ & $G_1$ & $G_2$ & p-value \\
\midrule
Work hard &	Reply number & \multirow{7}{*}{2.82\%$\pm$16.55\%} & 33.63\%$\pm$47.25\% & 12.42\%$\pm$32.98\% & $<1e^{-3}$ \\
Work hard &	Reply word count & & 36.37\%$\pm$48.11\% & \ 9.29\%$\pm$29.04\%  & $<1e^{-3}$ \\
Never miss &	Text similarity (DL) & & 23.47\%$\pm$42.39\% & 21.23\%$\pm$40.90\% & $<1e^{-3}$ \\
Never miss &	Text similarity (TF-IDF) & & 25.84\%$\pm$43.78\% & 16.59\%$\pm$37.21\% & $<1e^{-3}$ \\
Be polite &	Politeness & & 33.52\%$\pm$47.21\% & 13.54\%$\pm$34.21\% & $<1e^{-3}$ \\
Add references & Reference & & 28.49\%$\pm$45.14\% & 19.45\%$\pm$39.58\% & $<1e^{-3}$ \\
Make consensus & Mention other reviewer & & 28.92\%$\pm$45.35\% & 21.04\%$\pm$40.76\% & $<1e^{-3}$ \\
\bottomrule
\end{tabular}
}
\end{table}
In this section, we analyze the possible strategies to be employed during the rebuttal phase and the correlations between these strategies and outcomes.
We compile a list of rebuttal strategies for successful rebuttals from the literature, a set of review guidelines published by journals and conferences\footnote{https://iclr.cc/Conferences/2022/AuthorGuide}, and guidelines from experienced researchers\footnote{https://deviparikh.medium.com/how-we-write-rebuttals-dc84742fece1}.
First, based on our analysis in \secref{sec:rebuttal_results}, we define a review score increase (INC) after rebuttal as a \textbf{successful} rebuttal, while the rest of the rebuttals (\ie KEEP and DEC) are considered \textbf{non-successful}.
The overall success ratio is 17.95\%.
Second, we group the reviews for which authors did not submit any rebuttal as $G_0$ (the number in this group is 3,089).
From the authors' perspective, there are several approaches or strategies they can use to improve their submissions.
We have summarized the following quantifiable strategies:
\begin{itemize}
    \item \textbf{Work hard}: We use the number of authors' replies and the total word counts of authors' replies to investigate whether authors working harder (\ie more replies) will lead to an increased score. We define the number of author replies greater than 2 as $G_1$ and the remaining replies as $G_2$. For the word count, we define the top 33\% as $G_1$, and the bottom 33\% as $G_2$.
    \item \textbf{Be polite}: We use PoliteLex~\citep{li2020studying} to extract the positive and polite patterns in author responses, to verify whether being more polite will be helpful.
    We then rank the number of positive and polite patterns and divide the top 33\% into $G_1$ and the bottom 33\% into $G_2$. We can assume that $G_1$ is more polite than $G_2$.
    \item \textbf{Never miss}: To verify whether authors address all reviewers' concerns without missing any points, we use both TF-IDF text similarity and deep cosine similarity of sentence embeddings~\citep{cohan2020specter} to measure the similarity of reviews and author responses.
    We sort the similarity values and choose the top 33\% as $G_1$, and the bottom 33\% as $G_2$.
    We can assume that $G_1$ misses fewer concerns from reviewers than $G_2$.
    \item \textbf{Add references}: To determine whether authors respond with references, we use regular expressions to analyze the author's response to reviews and divide the references in the author's response into $G_1$, otherwise, it is $G_2$.
    \item \textbf{Make consensus}: Based on the SSNA in ~\secref{sec:ssna}, we compute whether the authors or reviewers mention other reviewers.
    As ICLR2022 provides reviewer IDs, we can easily count whether other reviewer IDs appear in the text or not\footnote{An example: https://openreview.net/forum?id=0IqFsR9wJvI\&noteId=hk92FJmSfz}.
    If the IDs are mentioned in the author responses, we group such author rebuttals as $G_1$, otherwise, it is $G_2$.
\end{itemize}

For each strategy, we divide the responded reviews into two groups and conduct analyses and t-tests on both groups to ascertain if the strategy leads to significant differences in the outcomes between the two groups.

The results are shown in ~\tableref{tab:strategy_analysis}.
We can find that:
\begin{inparaenum}[(1)]
\item Even if authors do not submit any rebuttals, reviewers may revise their review comments with a low probability (2.82\%).
\item The success rate of the group  that adopted the rebuttal strategy (\ie $G_1$) was significantly higher than that of the group that did not adopt the strategy. 
This suggests that when authors submit rebuttals, they need to be more skillful instead of simply replying.
\item The make consensus strategy is a particular example of a review process where the author's response can be enhanced by introducing peer effects.
\end{inparaenum}

\section{Rebuttal Success Prediction}
\label{sec:prediction}

In this section, we first define a \textbf{Rebuttal Success Prediction} task to validate our modeling on rebuttals.
Second, we present the possible features to model the successful/non-successful prediction task.
Lastly, we provide the results and analysis of our multi-factor model.
\subsection{Problem Definition}
For a review $r$ in paper $p$, we have the initial score as $s_0$.
After the authors complete the rebuttal, $r$ will receive the final score as $s_1$. Our task is to predict the sign for the final score $s_1 - s_0$ (\ie $s_1 - s_0 > 0$ or not).
We define the INC (\ie $s_1 - s_0 > 0$) as a successful rebuttal (1), and the other results (\ie KEEP and DEC) as non-successful (0).
It is an imbalanced binary classification task, so we use Macro-F1 and AUC to measure the performance of our models.
Macro-F1 is a variant of the F1-score, which is a harmonic mean of precision and recall.
AUC (Area Under the Curve) is a measure of how well a binary classification model can distinguish between positive and negative samples~\citep{pedregosa2011scikit}.
Both metrics are used to measure the performance of proposed model.

\subsection{Methods}

\begin{figure}
\centering
\includegraphics[width=0.8\linewidth]{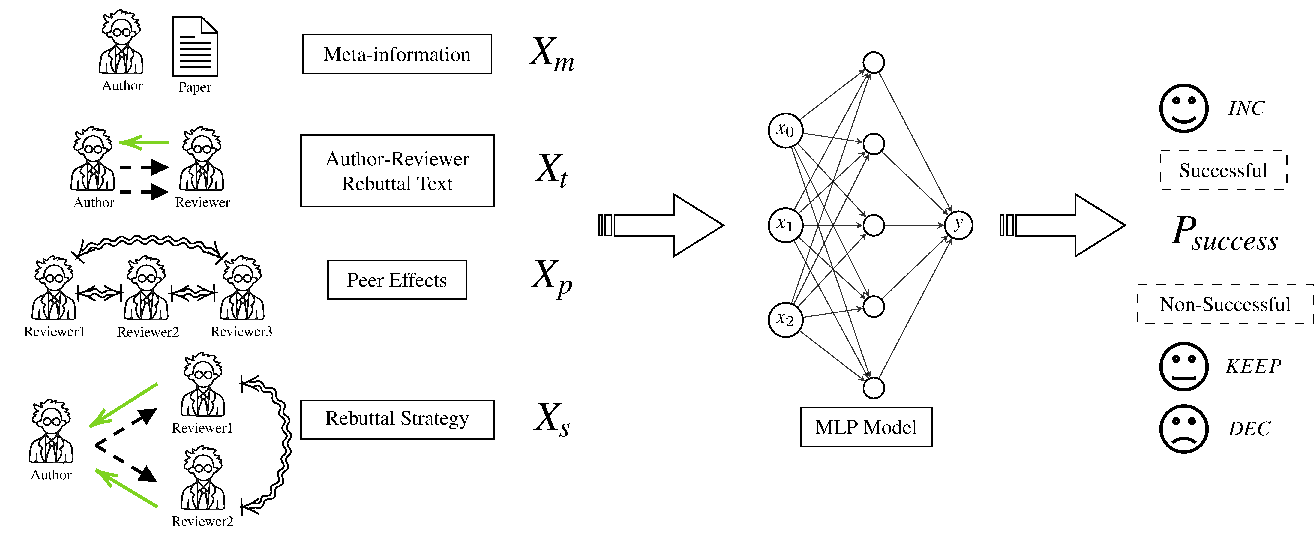}
\caption{The multi-factor prediction model for rebuttal success prediction.}
\label{fig:mlp_model}
\end{figure}

\begin{table}
\centering
\caption{Results of rebuttal success prediction.}
\label{tab:exp-results}
\scalebox{0.9}{
\begin{tabular}{ccccccccc}
\toprule
Model & Major Baseline & Random Baseline & MLP($X_m$) & MLP($X_t$) & MLP($X_s$) & MLP($X_p$) & MLP($X_p$, $X_m$, $X_s$, $X_t$)  \\
\midrule
AUC ($\uparrow$) & 0.5000 & 0.4898 & 0.6477 & 0.6285 & 0.7143 & 0.7704 & 0.7739  \\
\midrule
Macro-F1 ($\uparrow$) & 0.4379 & 0.4484 & 0.4379 & 0.4886 & 0.4605  & 0.6175 & 0.6401 \\
\bottomrule
\end{tabular}
}
\end{table}
Based on our previous discussion, we propose a multi-factor prediction model in \figref{fig:mlp_model} with the following input features:
\begin{itemize}
    \item Paper Meta-Information $X_{m}$: We use the meta-information of a paper $p$ in \tableref{tab:paper-analysis} to predict the sign of rebuttal, including 10 features (\ie $X_{m}\in \mathbb{R}^{N\times 10}$). 
    \item Rebuttal Text $X_{t}$: We use SPECTER~\citep{cohan2020specter} to encode both initial review text and author response text into 768-d dimension reviewer vectors $X_{t}^r \in \mathbb{R}^{N\times 768} $ and author vectors $X_t^a \in \mathbb{R}^{N\times 768}$. 
    \item Peer Effect Feature $X_{p}$: Following \cite{gao2019does}, we use the detailed scores of review $r$  (\ie Recommendation Score, Correctness, Technical Novelty and Significance, Empirical Novelty and Significance, and Confidence) and the  statistics (max/min/mean/median/std) of the detailed scores of other peer reviews  for a given paper $p$. 
    Besides, we add the number of balanced motifs and unbalanced motifs. 
    The peer effect features are  27-d dimension vectors (\ie $X_{p}\in \mathbb{R}^{N\times 27}$).
    \item Author Strategy Feature $X_{s}$: We use the strategy features employed in \tableref{tab:strategy_analysis} as the author strategy feature (\ie $ X_{s} \in \mathbb{R}^{N\times 7}$). 
\end{itemize}
After obtaining the above features, we concatenate these features, then encode these features through a two-layer  Multilayer Perceptron (MLP) model (the activation function is ReLU), and output the probability of success of the current review $r$ by a sigmoid function $f(x) ={\frac {1}{1+e^{-x}}}$:
\begin{equation}
\begin{aligned}
    X &= \mathrm{Concatenate}(X_m, X_t, X_p, X_s) \\
    P_{success} &= \mathrm{Sigmoid}(W_2 \cdot \mathrm{ReLU}(W_1 \cdot X+b_1) + b_2), \\
\end{aligned}
\end{equation}
where $W_1$, $b_1$, $W_2$ and $b_2$ are the parameters of MLP functions.
Besides, to verify the role of each feature, we input it separately as a baseline (\eg MLP($X_m$)).
Additionally, we compare two simple baselines: the majority baseline always chooses non-successful predictions, while the random baseline selects successful/non-successful predictions at random.
To validate the model, we perform train-validation-test splits of the data.
We exclude reviews whose authors have not received any responses.
Next, we randomly select 80\% of the reviews as training data and 10\% as validation data.
The remaining 10\% of reviews are used for testing model performance.
These models are implemented using PyTorch with the Adam optimizer (Learning Rate=0.01, Weight Decay=1e-3).
We choose Binary Cross Entropy (BCE) as our loss function and train for 1,000 epochs to select the model that performs best on the validation set, then report the performance on the test set.
The results are listed in \tableref{tab:exp-results}.

\subsection{Results}
From Table \ref{tab:exp-results}, we can observe the following:
\begin{inparaenum}[(1)]
\item Utilizing a machine learning model effectively improves prediction results (\ie AUC > 0.5).
\item Among the various feature types, the peer effect feature proves to be the most effective. This is consistent with previous findings ~\citep{gao2019does}.
\item The paper's meta-information feature exhibits poor performance, which can be attributed to a discrepancy between the paper's metadata and the rebuttal success prediction task.
\item Employing all features results in the best performance, surpassing that of other features.
\end{inparaenum}

\section{Conclusion}
\label{sec:conclusion}
\subsection{Summary and Discussion}
In this paper, we conduct an empirical study on the impact of a successful rebuttal stage in CS conference peer reviews. First, we collect and construct an open review dataset (ICLR2022) to examine the rebuttal stage at CS conferences.
Second, through a preliminary analysis of the dataset, we determine that the rebuttal stage is crucial for paper acceptance.
Third, we analyze the key factors for achieving a successful rebuttal, including reviewer social impacts and author rebuttal strategies.
We employ signed networks to investigate peer effects and discover that the balanced structure significantly increases after the rebuttal in all three top conference datasets.
Regarding author rebuttal strategies, we assess the effectiveness of several quantifiable approaches.
Finally, we develop a machine learning model to predict the success or failure of a review rebuttal in order to validate our findings.

We hope our research can illuminate strategies for crafting successful rebuttals for reviews and assist authors in getting their submissions accepted.
Undoubtedly, the most crucial aspect of a submission is the paper's quality.
Maintaining high quality requires diligent effort.
For instance, providing more details in appendices can strengthen the paper's foundation.
Submissions typically have strict upper limits for the main text, but allow unlimited pages for appendices and citations.
Enlisting the help of a skilled author is also recommended to improve submissions.
During the rebuttal stage, we employ social network analysis to evaluate changes in balanced network structures.
We observe an increase in the balanced network structure following the rebuttal stage.
This suggests that social pressure may play a critical role in influencing reviewers to modify their review scores.
Our findings align with previous studies~\citep{gao2019does}, but we offer an analytical perspective through social network analysis.
This can be utilized as a strategy (\eg building consensus) to encourage conformity and balance among reviewers.
Moreover, additional strategies (\eg being polite and remaining engaged) are advised to enhance the likelihood of a successful rebuttal.
These recommendations are also supported by related works on crafting detailed responses to reviewers in journal peer review~\citep{noble2017ten}.
Lastly, peer review platforms might consider refining the rebuttal process to make it more transparent and helpful, as the exchange between reviewers and authors can be viewed as an integral part of the scientific contribution to a paper.

\subsection{Limitation and Future Work}
There are several limitations to this study.
Regarding reviewers' social pressures, we only measured the interaction and influence among reviewers. However, interactions between reviewers and Area Chairs (ACs), which contribute to the final decision on a paper, are equally important. Unfortunately, due to data constraints, we were unable to assess the impact of this aspect.
In our examination of authors' rebuttal strategies, we employed the t-test hypothesis test, which only analyzes the mean variance of the data.
This statistical approach may have certain limitations, and additional experiments using causal analysis could be applied to assess strategies more effectively.
Moreover, in addition to the quantitative strategies mentioned in the paper, there are many strategies that are difficult to quantify.
Conducting research through questionnaires or interviews may also prove beneficial in exploring which strategies are most effective.

In terms of future work, we will concentrate on negative links (i.e., rejected decisions/recommendations) in peer review to identify the most significant (implicit/explicit) reasons behind these negative links.
Such insights can better enable authors to enhance their submissions and increase the likelihood of paper acceptance.
Moreover, comparing multiple conferences or analyzing a single conference over several years presents a valuable research direction. This approach allows us to assess the progress of peer review for conference organizers and contribute to the improvement of peer evaluation within the scientific community.

\section*{Acknowledgments}
This work is funded by the National Natural Science Foundation of China under Grant Nos. U21B2046, 62272125, and the National Key R\&D Program of China (2020AAA0105200). Huawei Shen is also supported by Beijing Academy of Artificial Intelligence (BAAI).

\bibliographystyle{apalike}

\bibliography{refs}

\end{document}